# Intertwined effects of elastic deformation and damage on vortex pinning and $J_c$ degradation in polycrystalline superconductors

Qing-Yu Wang[1], Shuai Hu[2], You-He Zhou[3], Cun Xue[2,*]

[1]School of Aeronautics, Northwestern Polytechnical University, Xi'an 710072, China

[2]School of Mechanics, Civil Engineering and Architecture, Northwestern Polytechnical University, Xi'an 710072, China

[3]Department of Mechanics and Engineering Sciences, College of Civil Engineering and Mechanics, Lanzhou University, Lanzhou 730000, China

*xuecun@nwpu.edu.cn

## Abstract

The damage and the critical current density ($J_c$) degradation of polycrystalline superconductors induced by strain dramatically influence their performance in applications. Unfortunately, the state-of-the-art experimental techniques are unable to detect the damage of internal polycrystalline structures and the microscopic superconductivity in the presence of strain. We propose a groundbreaking multi-scale theoretical framework aimed at revealing the underlying physical mechanisms of the reversible and irreversible $J_c$ degradation induced by the strain through tackling the complex intertwined effects of elastic deformation and damage on the superconductivity of grain boundaries and the associated vortex pinning. The results are well validated by experimental measurements. Utilizing the benchmarked physical model, we demonstrate that the damage evolutions of polycrystalline superconductors in the presence of strain can be approximately estimated by means of the electromagnetic experiments on $J_c$. Furthermore, we also discuss the characteristics of damage and $J_c$ degradation of polycrystalline superconductors subjected to biaxial mechanical loads. The findings will pave the way to investigate the tunable vortex pinning and $J_c$ of superconductors by strain, and to develop a brand new electromagnetic method to manifest the damage of polycrystalline superconductors.

Mechanical loads can induce damages [1] and critical property degradations [2-5] of superconductors, seriously threatening applications of superconducting devices. The plastic deformation, filament breakage, and even catastrophic destruction of superconductors have been observed in numerous experiments [6-8]. For superconducting thin films, damage evolution can be directly observed by scanning electron microscope (SEM) or transmission electron microscope (TEM). More recently, Zhou et al [1] pioneered the application of magneto-optical imaging (MOI) to investigate strain-induced damage evolution in coated conductors at macroscopic level, shedding light on the origins of initial damage. However, for polycrystalline superconductors such as $Nb_3Sn$ wires, which consist of a copper matrix and thousands of filaments [9], damage within the internal polycrystalline structures remains undetectable with the state-of-the-art experimental techniques.

Moreover, due to the strain sensitivity of polycrystalline superconductors such as $Nb_3Sn$ [10-14], the typical loss of critical current ($I_c$) with intrinsic axial-strain $\varepsilon_a$=0.40% at 4.07 K is approximately 40% [15], thereby reducing the quench current threshold and posing challenges to its safe and stable operation [16, 17]. Revealing the underlying mechanisms of strain-induced $J_c$ degradation is crucial for optimizing the electromagnetic properties of superconducting polycrystals, thereby directly influencing the applications of corresponding magnets. At the macroscopic level, numerous models [18-24] have been developed to characterize the strain dependence of critical current density of $Nb_3Sn$. However, the interpretation of experimental results and proposed models remain, for a significant part, empirical, and lacking a robust connection to the underlying physics. At the microscopic level, Godeke et al [25] revealed the strain sensitivity of $Nb_3Sn$ from the perspective of sub-lattice distortions. Note that key electromagnetic properties of superconductors are governed by thermal fluctuations, vortex-vortex interactions and the interaction between the pinning forces and quantized magnetic vortices [26-38]. Lipavsky et al [39] studied the effects of strain from the perspective of Ginzburg-Landau theory. In polycrystalline superconductors such as $Nb_3Sn$ and YBCO, grain boundaries (GBs) have been found to play a crucial role in vortex pinning [40-42]. van der Laan et al [43] investigated the effects of strain and GB angle on the critical current density of YBCO GBs. Building on the strain energy of dislocation, Yue et al [44] explored the mechanisms of current transport in the [001]-tilt low-angle GBs, yielding results consistent with the formula proposed by van der Laan et al. However, these studies on strain-induced $J_c$ degradation of polycrystalline superconductors focus on the reversible stage and have not been associated with vortex behaviors. The effects of strain on vortex pinning remains unclear and the mechanisms underlying the irreversible $J_c$ degradation of polycrystalline superconductors induced by applied strain are yet to be fully elucidated.

In this paper, a multi-scale computational framework, that integrates a multi-scale mechanical model with a physical model, is introduced. Through a 'reverse engineering' method to experimental $J_c$, we reveal the impact of strain on the superconductivity of grain boundaries and the associated vortex pinning in polycrystalline superconductors. Addtionally, without mechanical simulations, the internal damage can be approximated based on the electromagnetic experiments on $J_c$. The framework is further applied to analyze the $J_c$ degradation of polycrystalline superconductors under biaxial loads.

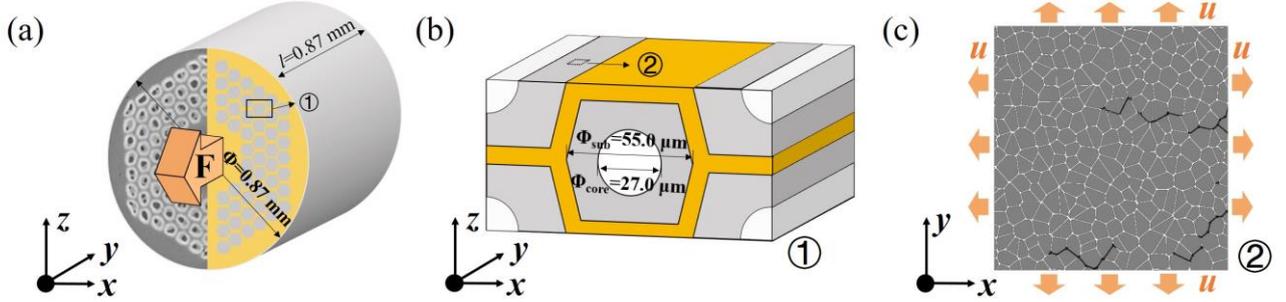

Fig. 1. (a) Schematic of the macroscopic model, where a IT $Nb_3Sn$ wire exposed to a tensile mechanical load. (b) Schematic of the mesoscopic model, consists of several sub-elements and copper matrix. (c) Schematic of microscopic model, where a polycrystalline $Nb_3Sn$ is subjected to displacements $u$ from the mesoscopic model.

The uniaxial quasi-static tensile tests are the most common experimental methods used to investigate the $J_c$ degradation of polycrystalline superconductors such as $Nb_3Sn$ wires. As shown in Fig. 1, in order to reproduce the mechanical response of $Nb_3Sn$ wire, we employ a numerical model consisting of an internal tin (IT) $Nb_3Sn$ wire. This wire, with a length of 0.87 mm and a diameter of 0.87 mm, comprises a copper matrix and 108 sub-elements and are subjected to tensile strain along the $y$-axis at a constant strain rate of $1\times10^{-5}$/s. To capture damage evolution in polycrystalline $Nb_3Sn$, we develop a multi-scale framework consisting of macroscopic, mesoscopic and microscopic models. The mesoscopic model (Fig. 1(b)) consists of copper matrix and sub-elements with an average size of 55 μm, while the microscopic model (Fig. 1(c)) consists of a 2D polycrystalline $Nb_3Sn$ systems with an area of $1.8\times1.8$ μm$^2$, featuring an average grain size of 108 nm. The underlying idea of the multiscale model is that taking local displacements from higher-level models as inputs to analyze the mechanical response of lower-level models. By employing this method, we can simulate the mechanical behaviors of which ranges from polycrystalline $Nb_3Sn$ to wire with acceptable computational resource.

To ensure the size of microscopic model is sufficiently large to investigate the mechanical behavior and damage evolution of polycrystalline $Nb_3Sn$, four randomly selected polycrystalline $Nb_3Sn$ (Panel 1 in Figs. 2(a-d)), located in different sub-elements, are analyzed. Panels 2-5 in Figs. 2(a-d) illustrate the damage evolution of these four samples with increasing applied strain. Previous experiments have demonstrated that the bare $Nb_3Sn$ filament exhibits brittle fracture with applied loads [45]. However, for the fiber-matrix composite structure of $Nb_3Sn$ wire, our simulation illustrates that polycrystalline $Nb_3Sn$ exhibits progressive damage. It may be attributed to that the copper matrix with higher toughness limits the brittle fracture of $Nb_3Sn$. Furthermore, as shown in Figs. 2(a-d), one can only observe intergranular fractures as the fracture strength of GBs is significantly lower than that of single grain [46].

Fig. 2(e) illustrates the elastic mechanical response of polycrystalline $Nb_3Sn$ at the beginning stage of applying strain. However, with further increasing applied strain $\varepsilon_a > 0.37\%$, local stresses exceed the fracture strength of GBs, resulting in crack initiation and propagation along GBs (Panels 2-5 in Figs. 2(a-d)). Four samples exhibit highly similar characteristics of the failure ratio of GBs, indicating that the simulated region is sufficiently large to eliminate the randomness of GB landscapes. In order to clearly capture the damage characteristics of polycrystalline $Nb_3Sn$, Fig. 2(f) presents the statistics of crack orientations at $\varepsilon_a = 0.64\%$, where $\theta$ represents the included angle between the crack orientation and the $x$-axis. It can be clearly seen that cracks are most likely to propagate along GBs with small $\theta$.

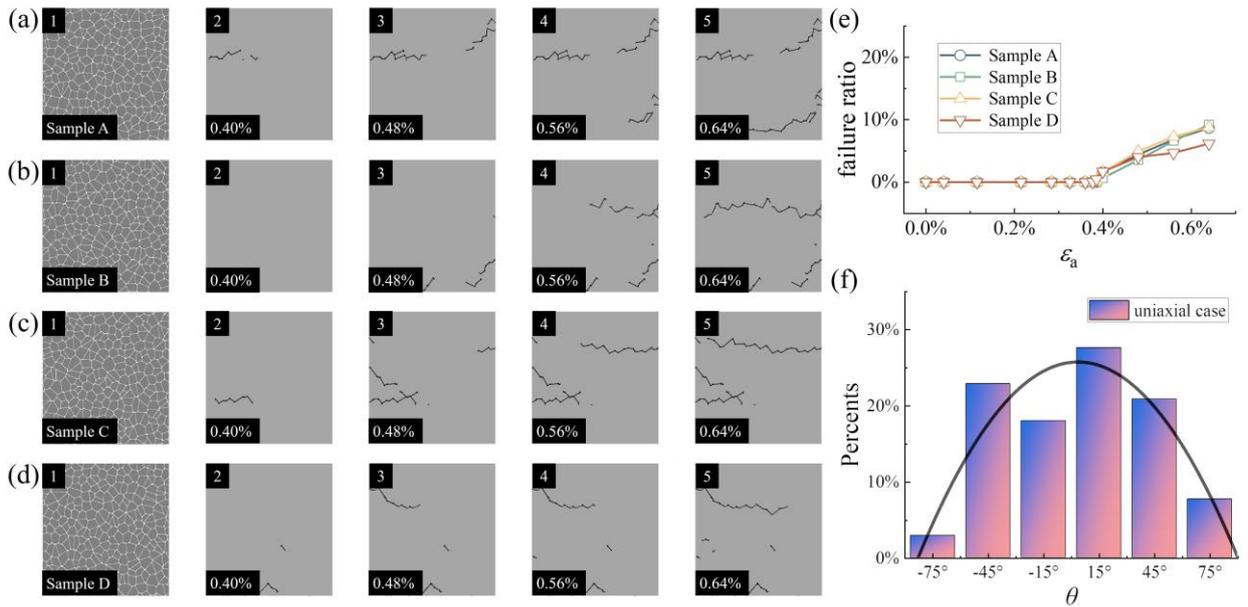

Fig. 2. (a-d) Four randomly selected samples and associated damage evolution with increasing applied strain ($\varepsilon_a = 0.40\% \rightarrow 0.48\% \rightarrow 0.56\% \rightarrow 0.64\%$). (e) The failure ratio of GBs in four samples of polycrystalline $Nb_3Sn$ versus strain. (f) Statistics of included angles at $\varepsilon_a = 0.64\%$. Black line represents the fitting curve for the statistics results.

Typical experimental $I_c$-$\varepsilon_a$ curves [15] clearly show two regions with different physical scenarios. One is the 'elastic' stage that the strain-induced $J_c$ degradation is reversible and can recover to initial state after fully unloading. Another is the 'plastic' stage where $J_c$ exhibits irreversible degradation. It is well known that the electromagnetic properties of superconductors are determined by vortex behaviors [47-54] like vortex-vortex interactions and the interaction between vortices and pinning force. For polycrystalline Nb$_3$Sn, its pinning effect is primarily resulted from the suppressed superconductivity on GBs [9]. Experiments [55, 56] have demonstrated that stress/strain plays a significant role in Cooper pair formation at pinning center. The observed similarities in $J_c$ attenuation and mechanical response of polycrystalline Nb$_3$Sn with applied strain inspire us to reveal the underlying physical mechanism of strain-induced $J_c$ degradation from the perspective of impact of strain on superconductivity of GBs and vortex pinning.

Time-dependent Ginzburg-Landau (TDGL) equations provides a powerful method to investigate the magnetic vortex behaviors[39, 57-61]. We propose a physical model to address the aforementioned problem based on TDGL theory [62-65],

$$\frac{u}{\sqrt{1+\gamma^2|\psi|^2}}\left(\frac{\partial}{\partial t}+\frac{\gamma^2}{2}\frac{\partial|\psi|^2}{\partial t}\right)\psi = (\nabla-i\mathbf{A})^2\psi+\left[1-T-g(\mathbf{r})-|\psi|^2\right]\psi, \tag{1}$$

$$\frac{\partial \mathbf{A}}{\partial t} = \text{Re}\left[\psi^*(-i\nabla-\mathbf{A})\psi\right]-\kappa^2\nabla\times\nabla\times\mathbf{A}. \tag{2}$$

where $\psi$, A and $\kappa$ represent order parameter, vector potential and GL parameter, respectively. $g(\mathbf{r})$ describes the pinning landscape of GBs, which is non-zero at GBs and remains zero inside grains. Note that $u = \pi^4/14\zeta(3) \approx 5.79$, and the parameter $\gamma=10$ accounts for inelastic scattering of the superconductor. In our simulations, the polycrystalline Nb$_3$Sn with an area of 1.8×1.8 μm$^2$ is exposed to a magnetic field $H_a$, a transport current $I_a$ and ambient temperature $T$. We use the magnetic field boundary conditions on the left and right borders, and periodic boundary conditions along the $y$-axis. The transport current along the $y$-axis is applied via the field boundary condition $H=H_a\pm H_I$, where $H_I$ represents the magnetic field induced by the current $I_a$. To efficiently simulate the vortex behaviors in the polycrystalline Nb$_3$Sn, a stable implicit numerical algorithm for TDGL theory implemented on GPU [66] is introduced in Ref. [67].

Experiments indicated that with different mechanical loading modes, invariant strain function [21, 68] gives a more unified analytical treatment of strain dependence of critical properties of Nb$_3$Sn tapes in reversible regions. In order to obtain a general formula for the impact of strain on superconductivity of GBs, we employ a function with an independent variable of the second invariant of strain tensor $J_2$. In our physical model, investigating the effects of strain on $J_c$ requires the superconductivity of GBs at zero strain state ($|\psi_{GB0}|$).

However, to date, the value of $|\psi_{GB0}|$ of polycrystalline $Nb_3Sn$ measured by experiments has not been reported. Therefore, we conduct TDGL simulations using a series of $|\psi_{GB0}|$ (0.90, 0.80, 0.75, 0.65, 0.60, 0.50 and 0.40). In the following text, we take one $|\psi_{GB0}|$ to illustrate our physical model.

For the case with $|\psi_{GB0}|=0.75$, the impact of strain on $|\psi_{GB}|$ is obtained through the 'reverse engineering' method by tuning $|\psi_{GB}|$ to reproduce the experimental data of $J_c$-$\varepsilon_a$. Fig. 3(a) illustrates the experimental $J_c$-$\varepsilon_a$ and associated simulated results. The critical current density decreases monotonically with increasing applied strain. In the 'elastic' stage of polycrystalline $Nb_3Sn$, where no cracks initiate, the degradation of $J_c$ is solely determined by strain-induced suppression of $|\psi_{GB}|$ (points A, B, C and D in Fig. 3(a)). This suppression is reversible and can recover to its initial state after unloading. Once $\varepsilon_a$ exceeds $\varepsilon_{irr}$, the blue line illustrates the $J_c$ degradation induced by reversible strain and cracks (points E, F, G and H in Figs. 3(a)). In contrast, if we only consider the effect of reversible strain on $|\psi_{GB}|$, corresponding $J_c$ attenuation of $Nb_3Sn$ is shown by the yellow line. In the irreversible stage, there exist significant differences between experiments and the simulated results without considering damage. It indicates that the damage of polycrystalline $Nb_3Sn$ gradually dominates in $J_c$ degradation with further increasing $\varepsilon_a$. Besides, the strain-induced cracks are not recoverable, resulting in the irreversible degradation of $J_c$. Surprisingly, the simulated $J_c$ after partial unloading (points E', F', G' and H' in Figs. 3(a)) is in good agreements with corresponding experimental results, providing clear evidence for the validation of the multi-physics computational framework.

The results of cases with other $|\psi_{GB0}|$ can be obtained utilizing a similar procedure as $|\psi_{GB0}|=0.75$. Fig. 3(b) illustrates the experimental $J_c$-$\varepsilon_a$ and simulated $J_c$ degradation with $|\psi_{GB0}|=0.40$. For cases where $|\psi_{GB0}|\geq0.80$, the simulated results indicate that $J_c$-$\varepsilon_a$ exhibits non-monotonicity with tensile strain, which is not consistent with experimental results [15] (further details available in the SI). For the case with $|\psi_{GB0}|\leq0.40$, there exist deviations between experimental and simulated $J_c$ with increasing applied strain, suggesting that small $|\psi_{GB0}|$ cannot accurately reproduce the experimental results [15]. Another evidence [69] also demonstrated that the simulation with $|\psi_{GB0}|\leq0.40$ cannot reproduce the experimental $J_c$ of $Nb_3Sn$ well, where $J_c$ is enhanced with increasing grain size at high fields. Therefore, the reasonable value of $|\psi_{GB0}|$ should be in the scopes from 0.40 to 0.75. Fig. 3(c) represents the $|\psi_{GB}|$-$J_2^{0.5}$ curves for $|\psi_{GB0}|$ of 0.75, 0.60 and 0.40, obtained through the 'reverse engineering' method. The superconductivity of GBs decrease with increasing applied strain, and three $|\psi_{GB}|$-$J_2^{0.5}$ curves exhibit similar trends.

For a given applied current, Panel 3 of Figs. 3(e) shows that no continuous vortex motion occurs in the absence of applied loads, indicating that the applied Lorentz force is insufficiently large to move vortices along the GBs or drag vortices out of the GBs. However, as applied strain is increased to 0.13%, the vortex motion is observed, especially along the GBs with an included angle close to 0° (Panels 4-6 of Fig. 3(e)). It is attributed to the fact that the strain suppresses the Cooper pair formation at the GBs, thereby reducing the pinning effects. As $\varepsilon_a$ exceeds the irreversible strain, cracks initiate and propagate along the GBs. Panels 7-8 of Fig. 3(e) illustrate that cracks act as flux flow channels and promote the vortex motion.

As a matter of fact, our physical model enables the approximation of damage in superconducting polycrystals based on the electromagnetic response and thus the mechanical simulations are not necessary, particularly for the internal damage of wires that cannot be directly observed by SEM experiments. For a $Nb_3Sn$ wire, the $J_c$ degradation with a given stress/strain state can be obtained by transport measurements. Simultaneously, numerical simulations for polycrystalline $Nb_3Sn$ without considering damage are carried out to determine the critical current density. As illustrated in Fig. 3(d), the difference between two results roughly reflects the damage evolution of $Nb_3Sn$ wire, which increase with applied strain and exhibits similar characteristics with the failure ratio of GBs shown in Fig. 2(e).

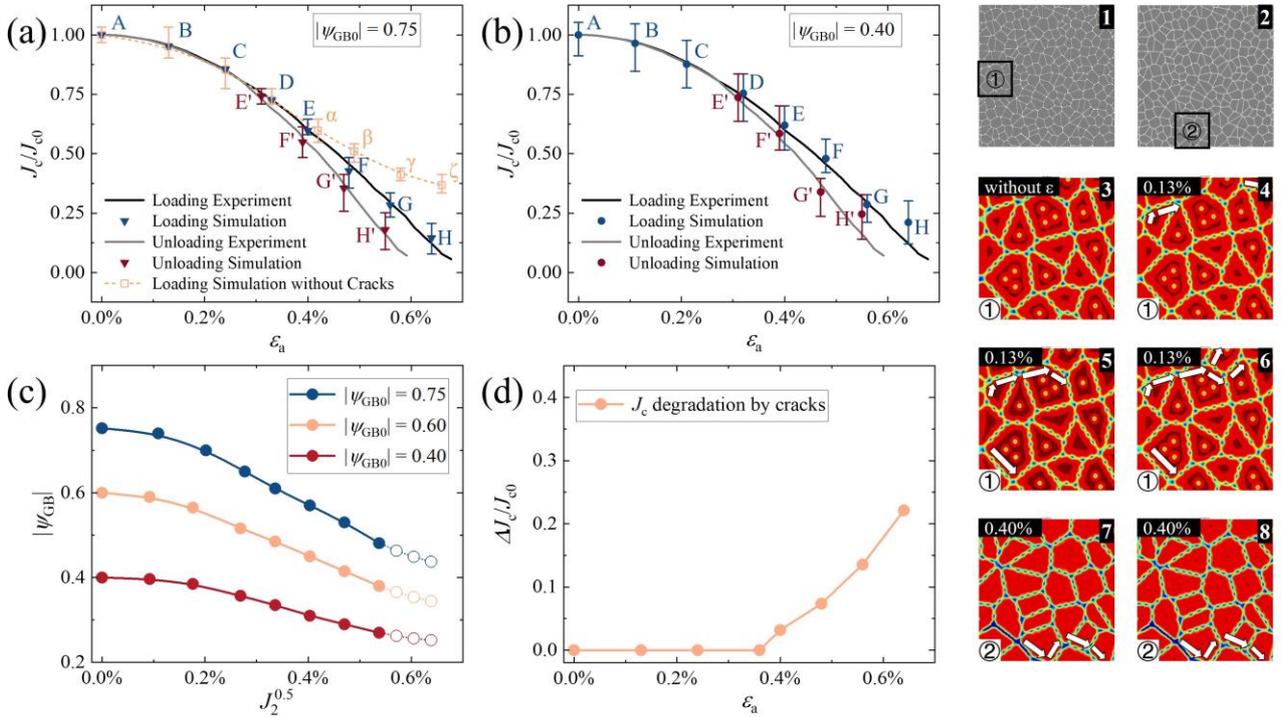

Fig. 3. (a-b) Black solid line and gray solid line represent the experimentally measured $J_c$-$\varepsilon_a$ under loading and partial unloading, respectively. Blue solid symbols and red solid symbols represent the average simulated $J_c$ for four samples under loading and partial unloading. The yellow empty symbols in (a) represent the average simulated $J_c$ and without consideration of cracks. (c) $|\psi_{GB}|$ as a function of $J_2^{0.5}$ with different $|\psi_{GB0}|$(0.75, 0.60 and 0.40). Dashed lines represent the extended data. (d) $J_c$ degradation caused by cracks. (e) Panels 1-2 refer to the damage of Sample B without and with damage. Panels 3-8 indicate the corresponding vortex behaviors at different moments. The white arrows represent the paths of vortex motion.

Through the experimental $J_c$ degradation under uniaxial tensile loads, we confirm the validation of the proposed computational framework. Unlike the uniaxial loading condition, superconducting magnets are subjected to complicated Lorentz force during the current ramping process. Accurately understanding the $J_c$ degradation under complicated loads is crucial for the property evaluation of superconductors and stable operation of electromagnets. In the following text, we explore the characteristics of mechanical behaviors and strain dependent $J_c$ of Nb$_3$Sn wire for biaxial case.

A representative volume element (RVE) of a typical Nb$_3$Sn magnet is considered. Fig. 4(a) illustrates the RVE with a length of 0.95 mm and a cross-sectional area of 0.95×0.95 mm$^2$, consisting of an IT Nb$_3$Sn wire and surrounding epoxy. The RVE is subjected to a biaxial load, i.e., an applied compressive strain along the $x$-axis and an applied tensile strain along the $y$-axis. Fig. 4(b) illustrates the average maximum principal stress $\sigma_a^p$ and the maximum principal stress $\sigma_{max}^p$ versus $J_2^{0.5}$. $\sigma_a^p$ increase linearly with the $J_2^{0.5}$ and reaches a peak of 164 MPa. The damage initiates at $J_2^{0.5}$=0.33% and the maximum principal stress exists intensive jumps occur as $J_2^{0.5}$ further increases (subgraph in Fig. 4(b)), attributed to the concentrated crack initiations and propagations. Fig. 4(c) demonstrates the damage evolution of polycrystalline Nb$_3$Sn with biaxial loads at $J_2^{0.5}$=0.33%, 0.37%, 0.46%, and 0.52%. Unlike the uniaxial tensile cases, cracks tend to orient at $\theta = \pm 45°$ with $J_2^{0.5}$=0.52% (Panels 1-5 of Fig. 4). It indicates that the mechanical behavior and damage evolution of polycrystalline Nb$_3$Sn depend strongly on the mechanical loading modes.

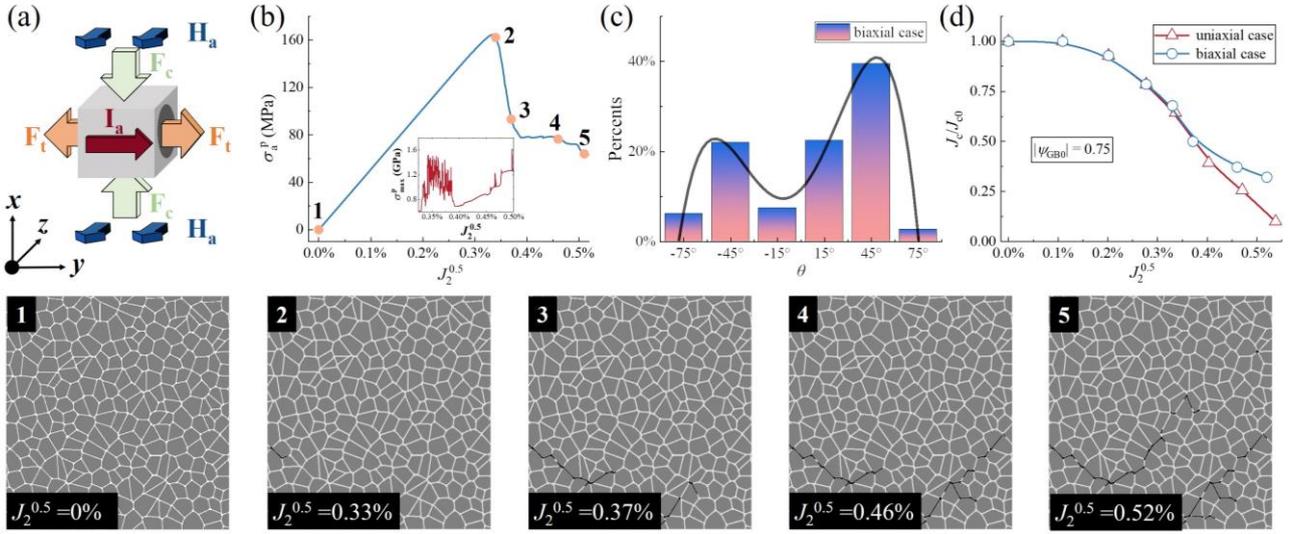

Fig. 4. (a) Schematic of a RVE of Nb$_3$Sn magnet exposed to a transport current $I_a$, an external magnetic field $H_a$, and a biaxial mechanical load. (b) the average maximum principal stress $\sigma_a^p$ (blue line) and the maximum principal stress $\sigma_{max}^p$ (red line) versus $J_2^{0.5}$. (c) The statistics of crack orientations at $J_2^{0.5}$=0.52%. Black line represents the fitting function for the statistics results. (d) The simulated $J_c$-$J_2^{0.5}$ curves for uniaxial (red line) and biaxial (blue line) cases. (e) Damage evolutions of polycrystalline Nb$_3$Sn with different strains ($J_2^{0.5}$=0%→0.33%→0.37 %→0.46%→0.52%).

Fig. 4(d) illustrates the strain dependence of critical current density for biaxial and uniaxial cases. In the reversible stage ($J_2^{0.5}$<0.33%), the $J_c$ degradation under biaxial loads show good agreements with that under uniaxial tensile loads. The consistent degradation of critical properties versus $J_2^{0.5}$ under different mechanical loading conditions is also reported in Ref. [68]. However, as the strain exceeds the irreversible strain, there exists obvious difference in $J_c$ degradation between uniaxial and biaxial cases, suggesting $J_c$ attenuation is strongly dependent on the mechanical loading modes. It is attributed to the fact that the significant difference of crack orientations leads to distinct critical vortex depinning of vortices.

In conclusion, we proposed a multi-scale computational framework integrating a multi-scale mechanical model with a physical model. The validation of the computational framework is confirmed through the experimental $J_c$ degradation with uniaxial tensile loads. The impact of strain on superconductivity and vortex pinning of polycrystalline superconductors is revealed. The irreversible $J_c$ degradation of polycrystalline superconductors is mainly attributed to the suppression of vortex pinning induced by both reversible strain and irreversible damage. The most striking finding is that, without mechanical simulations, the proposed physical model can be utilized to approximate the internal damage of superconducting polycrystals based on the electromagnetic experiments on $J_c$. Finally, we explore the $J_c$ degradation of polycrystalline superconductors subjected to biaxial loads by the computational framework. Our findings indicate that the irreversible $J_c$ degradation is strongly dependent on the mechanical loading modes. The computational framework paves the way for mechanical tuning superconductivity and associated vortex pinning, and exploring the damage of superconducting polycrystals through electromagnetic responses.

*Acknowledgements*—We acknowledge support by the National Natural Science Foundation of China (Grant No. 12372210). We also acknowledge support by the Fundamental Research Funds for the Central Universities lzujbky-2024-jdzx02. We thank for the helpful discussions in mechanical analysis of polycrystalline structures by He Ding at North university of China.